\documentclass[
 aip,
 apl,
 sd,
 amsmath,amssymb,
 reprint,
]{revtex4-1}

\usepackage{graphicx}% Include figure files
\usepackage{float}
\usepackage{dcolumn}% Align table columns on decimal point
\usepackage{bm}% bold math
\usepackage{amsmath}
\usepackage[english]{babel}
\usepackage{braket}
\usepackage{epstopdf}
\usepackage{siunitx}

\begin{document}

\preprint{AIP/123-QED}

\title{Double-sided coaxial circuit QED with out-of-plane wiring}

\author{J. Rahamim}
\author{T. Behrle}
\author{M. J. Peterer}
\author{A. Patterson}
\author{P.\,A. Spring}
\author{T. Tsunoda}
\author{R. Manenti}
\author{G. Tancredi}
\author{P.\,J. Leek}
\affiliation{ 
Clarendon Laboratory, Department of Physics, University of Oxford, Parks Road, Oxford OX1 3PU, United Kingdom
}

\date{\today}

\begin{abstract}
Superconducting circuits are well established as a strong candidate platform for the development of quantum computing. In order to advance to a practically useful level, architectures are needed which combine arrays of many qubits with selective qubit control and readout, without compromising on coherence. Here we present a coaxial circuit QED architecture in which qubit and resonator are fabricated on opposing sides of a single chip, and control and readout wiring are provided by coaxial wiring running perpendicular to the chip plane. We present characterization measurements of a fabricated device in good agreement with simulated parameters and demonstrating energy relaxation and dephasing times of  $T_{1}\,=\,4.1\,\mu$s and $T_{2}\,=\,5.7\,\mu$s respectively. The architecture allows for scaling to large arrays of selectively controlled and measured qubits with the advantage of all wiring being out of the plane.
\end{abstract}


\maketitle

The realization of technological devices that harness quantum superposition and entanglement to perform computational tasks that are difficult with classical computers is a major research goal that may revolutionize computing\cite{Ladd:2010aa}. Superconducting circuits have advanced to become a strong candidate platform for building such quantum computers\cite{Schoelkopf2013}, with recent demonstrations of circuit operation at the threshold for fault tolerance\cite{Barends:2014aa}, quantum error detection\cite{Riste2015,Kelly2015} and correction\cite{Ofek:2016aa}, and rudimentary quantum simulations\cite{Salathe2015,Barends2015,OMalley2016}. While the scale required for full fault tolerant universal quantum computation is still far away\cite{Fowler:2012aa}, current devices are not far from the complexity required for a demonstration of computation that is beyond the reach of the best classical supercomputers\cite{Boixo:2016aa}. To reach beyond this scale (of order 50 qubits) in a single monolithic quantum circuit, it is desirable to develop circuit architectures that implement good connectivity among arrays of many qubits, along with selective control and readout wiring, without compromising on qubit coherence. This is difficult to achieve if the circuit is constrained to a single 2D plane, since the number of control and readout connections scales linearly with the number of qubits $N$, while the edges of a 2D array scale as $\sqrt{N}$. This problem can only be overcome by incorporating 3D connectivity.

The challenge of incorporating control wiring out of the plane of a superconducting quantum circuit has been approached so far from several directions. 
A recent proposal suggests the use of through-chip microwave silicon vias, as part of a monolithic architecture to implement the surface code \cite{Versluis2016}. Bump bonding between multiple circuit layers \cite{MutusMarch2017}, and spring-loaded microwave contacts \cite{Bejanin2016} are also under development. Pursuing a modular (as opposed to monolithic) quantum computing architecture is an alternative route, and some promising steps have been made in this direction with superconducting circuits, through integration with high quality 3D microwave resonators \cite{Axline:2016aa,Brecht:2016aa,Narla2016}. 

In this letter we present a single unit cell of an architecture for quantum computing with superconducting circuits that is simple to fabricate, requires no bonds, exploits only capacitive couplings, and implements qubit control and readout entirely out of the plane of the qubit, without relying on complex through-chip fabrication\cite{Tolpygo:2016aa}.

\begin{figure}[b!]
\includegraphics[width=\linewidth,trim=0cm 0.5cm 0cm 0.5cm]{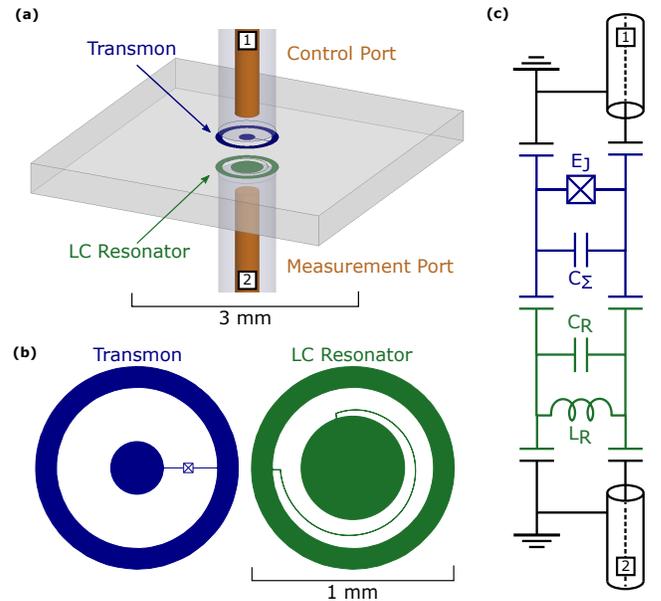}
\caption{\label{fig:device} 
(a) CAD design of the unit cell, with transmon qubit and lumped element resonator on opposing sides of a substrate, and control and measurement ports perpendicular to the chip plane. (b) Designs of the transmon and resonator. In the transmon the two electrodes are connected by a single Josephson junction, whereas the electrodes of the resonator are connected by an inductor line. (c) Equivalent circuit of the device, showing the resonator inductance and capacitance, $L_{R}$ and $C_{R}$, the junction Josephson energy $E_{J}$ and effective capacitance over the junction $C_{\Sigma}$.}
\end{figure} 

By virtue of the out-of-plane readout and wiring elements the device may be physically scaled to large 2D qubit arrays without any alteration to the wiring design. Additionally the double-sided structure and absence of wiring elements in the circuit design avoids crowding on the chip, hence reducing sources of crosstalk.

The device is depicted in Fig.~\ref{fig:device}. It consists of a superconducting charge qubit in the transmon regime\cite{Koch:2007aa} with coaxial electrodes, which we call the coaxmon (similar to the concentric \cite{Braumuller:2016aa} and aperture\cite{Brecht:2016aa} transmons) coupled to a lumped element LC microwave resonator fabricated on the opposite side of the chip, realising dispersive circuit quantum electrodynamics\cite{Wallraff2004} (QED). The device is controlled and measured via coaxial ports, perpendicular to the plane of the chip (see Fig.~\ref{fig:device} (a)), whose distance from the chip can be modified to change the external quality factor of the circuits. These ports can be used for independent control of the qubit and measurement of the resonator in reflection, or to measure the device in transmission.

The device is fabricated through two stages of electron beam lithography, patterning either side of a $0.5\,$mm thick sapphire chip with an aluminum LC resonator and coaxmon. During fabrication the bottom of the chip is protected with a spin-coated layer of polymer resist, and chip holders are used to ensure the bottom of the device is suspended throughout. The process could be further improved in the future by producing the LC resonators with photolithography thus enabling batch production of devices that only require one electron-beam step. The device is then mounted in an aluminum sample holder and thermally anchored to the  $10\,$mK base plate of a dilution refrigerator. The control and measurement ports consist of copper-beryllium wire passing through a cylindrical hole in the sample holder, soldered to the center conductor of a microwave connector in order to connect to external microwave wiring. In this experiment the distance from the qubit(resonator) to the control(measurement) port is $0.6(0.4)\,$mm. The device is embedded in a standard circuit QED measurement setup, in which input signals are heavily cryogenically attenuated (by approximately $70\,$dB) to reduce thermal noise, and measurements are made via cryogenic circulators and a low noise HEMT amplifier, the signal finally being recorded as a voltage $V_{ADC}$ with an analog-to-digital converter (ADC).

\begin{figure}[t]
\includegraphics[width=\linewidth,trim=0cm 0.5cm 0cm 0.5cm]{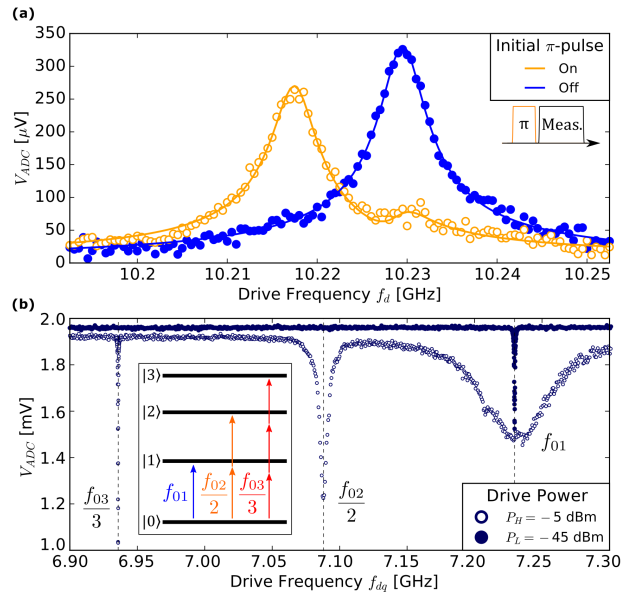}
\caption{\label{fig:cqed} 
(a) Resonator spectroscopy in the low photon number limit $\bar{n}\ll1$. Transmitted signal amplitude at the ADC measured with a $1\,\mu$s pulse at $f_{d}$, with (orange) and without (blue) a $\pi$-pulse applied to the qubit immediately prior to the measurement pulse (pulse scheme inset). The data (circles) are fitted (solid lines) as described in the main text. (b) Pulsed qubit spectroscopy for two different qubit drive powers. At $-45\,$dBm only the $f_{01}=7.23\,$GHz transition is visible. At a drive of $-5\,$dBm, two multi-photon transition frequencies $f_{02}/2$ and $f_{03}/3$ become visible and the $f_{01}$ is broadened. Inset: the energy level diagram of a transmon qubit illustrating the multi-photon transitions.}
\end{figure}

We first measure the device transmission spectrum $S_{21}$ at a low drive power of $P_{r}\,=\,-50\,$dBm, finding the Lorentzian response of the LC resonator\cite{sampleholder} at $f_{r0}=10.23\,$GHz, with quality factor $Q=2080$ (see Fig.~\ref{fig:cqed} (a)). Far from resonance, $S_{21}$ remains $30\,$dB below the LC resonance over the $8-12\,$GHz measurement bandwidth. We next fix the measurement drive at the LC resonance, and add an additional drive at frequency $f_{dq}$ to port 1, to carry out spectroscopy of the qubit using the dispersive qubit state-dependent frequency shift of the LC resonator \cite{Wallraff2005}. The spectroscopy is carried out with an $8\,\mu$s drive pulse immediately followed by an $8\,\mu$s measurement pulse at frequency $f_{r0}$ and power $P_r=-35\,$dBm, averaging the data $10^{6}$ times. In Fig.~\ref{fig:cqed} (b) we show such spectroscopy at two different drive powers. At low drive power $P_L=-45~\rm{dBm}$ we observe only the qubit transition at $f_{01}=7.23~\rm{GHz}$, whereas at higher power $P_H=-5~\rm{dBm}$ we observe two additional spectral lines below $f_{01}$, as expected of a transmon qubit. We observe a two-photon transition at  $f_{02}/2=7.08~\rm{GHz}$ and a three-photon transition at  $f_{03}/3=6.93~\rm{GHz}$ to higher energy levels of the transmon, as illustrated in the inset of  Fig.~\ref{fig:cqed} (b). Note that the broadening of the $f_{01}$ peak at the higher drive power originates from strong Rabi driving of the transition. From these parameters we calculate a detuning between qubit and resonator of $\Delta_{0}/(2\pi)\,=-3.00\,$GHz, Josephson energy $E_{J}/h\,=\,24.1\,$ GHz, charging energy $E_{C}/h\,=\,294\,$MHz, and  $E_{J}/E_{C}=81.8$.

We next characterize the interaction between qubit and resonator by measuring the qubit-state-dependent resonator frequency shift $\chi$. In order to do this, we repeat the transmission measurement of the LC resonance after preparing the qubit in its first excited state prior to a measurement pulse (see Fig. \ref{fig:cqed} (a) orange curve). The resonance is seen to shift from $f_{r0}$ to $f_{r1}=10.217\,$GHz. In addition to the shifted peak at $f_{r1}$, a residual peak at $f_{r0}$ is also visible due to the excited state population partly decaying during the measurement pulse. The response is fitted to the weighted sum of two Lorentzians in the complex plane, from which we extract the dispersive shift of the resonator $2\,\chi/(2\pi)\,=f_{r0}-f_{r1}=\,-12.68\,$MHz. We then use this to derive the qubit-resonator coupling  $g/(2\pi)\,=\,462\,$MHz from the relation
\begin{equation}
	\chi\approx-\frac{g^{2}(E_{C}/\hbar)}{\Delta_{0}(\Delta_{0}-(E_{C}/\hbar))},
\end{equation}

valid for a transmon in the dispersive regime\cite{Koch:2007aa}. Since our implementation of cQED consists entirely of lumped elements, we can calculate the expected parameters using a finite element electrostatic simulation (Ansys Maxwell) of the circuit. The circuit representation can be quantized to give expressions for the qubit and resonator frequencies, $f_{01}$ and $f_{r0}$, and the coupling $g$ between them, as a function of the capacitance network, as well as the resonator inductance $L_R$ and Josephson energy $E_J$ which we match to the experimentally measured values. Such a simulation predicts a coupling $g/2\pi\approx420\,$MHz. The discrepancy between the estimated and measured value may be due to the use of a static solver, which neglects any inductive coupling in the circuit. We have also used this model to simulate the coupling between control (measurement) port and qubit (resonator), and its dependence on the displacement of the port axis from the qubit and resonator centers. We find that for the circuit geometry presented here, the coupling falls to $\sim5\%$ at a displacement of 1 mm, indicating that good selectivity should be achievable between control and measurement signals in adjacent cells in a grid of multiple qubits.

\begin{figure}[t]
\includegraphics[width=\linewidth,trim=0cm 0.5cm 0cm 0.5cm]{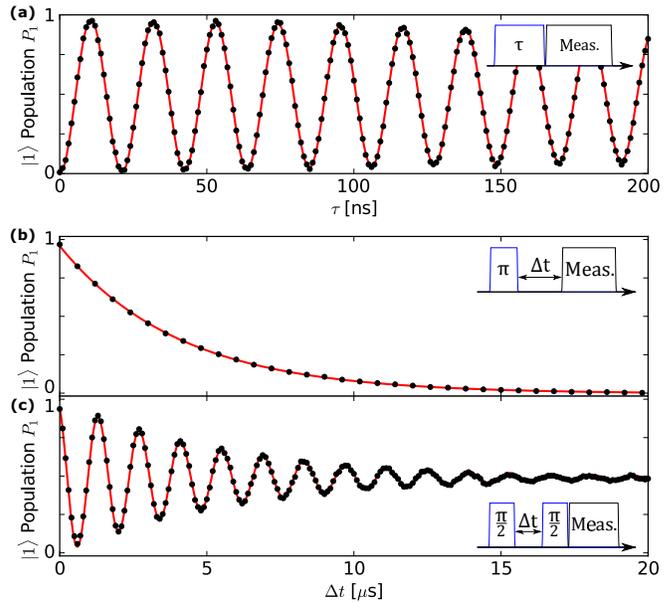}
\caption{\label{fig:coherence} 
Time-resolved qubit measurements with data points in black and fits as solid red lines. The pulse schemes of the measurements are shown in the insets. (a) Rabi oscillations of $47\,$MHz performed on the device at a qubit drive power of  $-20\,$dBm. An exponentially decaying oscillation  $P_{1}=A\cos(\omega t)e^{-t/T_{\rm{Rabi}}}$ is fitted to the data. (b) Qubit energy relaxation fitted to an exponential decay $P_{1}=Ae^{-t/T_{1}}$ reveals a $T_{1}\,=\,4.10\,\mu$s. (c) Ramsey oscillations performed with the qubit drive detuned $4.5\,$ MHz from $f_{01}$ reveals a $T_{2}\,=\,5.65\,\mu$s from the fitted oscillating decay  $P_{1}=A\cos(\omega t+\phi)e^{-t/T_{2}}$. In all cases $P_{1}$ is extracted from the pulsed resonator response by fitting to Cavity-Bloch equations. }
\end{figure}

We now move on to time resolved  qubit measurements which are performed by measuring the resonator in reflection on port 2 and applying qubit drive pulses to port 1. In Fig \ref{fig:coherence} (a) we first show Rabi oscillations of the qubit state, measured by first applying a short microwave pulse of length $\tau$ to the qubit in its ground state at frequency $f_{01}$, followed by a resonator readout pulse of length 16 $\mu$s and frequency $f_{r0}$ at a low photon number. The population $P_{1}$ of the qubit excited state $\ket{1}$ is recovered from the weighted integral of the resonator response by comparing it to the integral of simulated Cavity-Bloch traces\cite{Bianchetti2009} using parameters independently determined by the other characterization experiments, and including a correction to take into account interference with the directly reflected measurement pulse.

We determine the qubit relaxation time $T_{1}\,=\,4.10\,\mu$s and phase coherence time $T_{2}\,=\,5.65\,\mu$s using standard techniques (see Fig. \ref{fig:coherence} (b) and (c)). A spin echo pulse sequence reveals an extended $T_{2E}\,=\,6.67\,\mu$s. To further evaluate the performance of the device we perform Clifford-based randomized benchmarking and find the average fidelities of primitive gates to be 99.5\% using half-DRAG pulses\cite{Lucero2010}. We also determine an upper bound for the qubit temperature by measuring the amplitude of Rabi oscillations on the $f_{12}$ transition both with and without an initial $\pi$-pulse on the $f_{01}$ transition\cite{Geerlings2013}. We find the qubit temperature to be $T_q\le70\,$mK corresponding to an initial ground state population of $P_{0}\ge99.3\%$. Hence our single-qubit unit cell displays promising performance for an initial demonstration.

\begin{table}
\caption{ Device Parameters.}
\label{tab:QParams}
\setlength{\extrarowheight}{.3em}
\begin{tabular*}{\linewidth}{l@{\extracolsep{\fill}}c}
	\hline\hline
	Parameter		&	Value 	\\ \hline
	Resonator Frequency $f_{r0}$ [GHz]		&	10.23		\\ 
	Resonator Quality Factor			&	2080			\\
	Qubit $f_{01}$ [GHz]		&	7.23		\\ 
	Dispersive shift $\chi/2\pi$ [MHz]			&	-6.34 	\\
	$E_{C}/h$ [MHz]		&	294		\\
	$E_{J}/E_{C}$	&	81.8			\\ 
	Coupling $g/2\pi$ [MHz]			&	462	\\ 
	$T_{1}$ [$\mu$s]			&	4.10	\\
	$T_{2}$ [$\mu$s]			&	5.65		\\ 
	$T_{2}\,$Echo [$\mu$s]			&	6.67	\\ \hline\hline
\end{tabular*}
\end{table}

% PROSPECTS/CONCLUSIONS
We have presented a double-sided coaxial implementation of circuit QED. We summarize the device parameters in table \ref{tab:QParams}. We anticipate this architecture to be easily extendable to arrays of nearest-neighbor coupled qubits by virtue of the out-of-plane readout and control wiring, and so will be a good candidate architecture for the next generation of multi-qubit devices for quantum simulation and computation.

\section{Acknowledgments}

This work has received funding from the UK Engineering and Physical Sciences Research Council under grants EP/J001821/1, EP/J013501/1 and EP/M013243/1. AP and TT acknowledge Oxford Instruments Nanoscience and the Nakajima Foundation respectively for financial support. Aspects of this work are currently covered by an international patent application.


\begin{thebibliography}{25}
\bibitem []{sampleholder}%
  \BibitemOpen
  \bibfield  {author} {\bibinfo {author} {\bibfnamefont {In order to rule out the possibility that this resonance is a mode of the sample holder, we have performed high frequency simulations of the device predicting the LC resonance at $10.5\,$GHz and the lowest frequency mode of the sample holder at $20.5\,$GHz}}}
  \BibitemShut {NoStop}%
\end{thebibliography}
\end{document}